\documentclass[prl,twocolumn,amsmath,amssymb,floatfix]{revtex4}

\usepackage{graphicx}
\usepackage{bm}

\begin{document}

\title{Magnetic Resonance Force Microscopy of paramagnetic electron spins at millikelvin temperatures}

\author{A. Vinante}
\email{vinante@physics.leidenuniv.nl}

\author{G. Wijts}
\author{O. Usenko}
\author{L. Schinkelshoek}
\author{T.H. Oosterkamp}
\affiliation{Leiden Institute of Physics, Leiden University, the Netherlands} 

\date{\today}

\begin{abstract}
Magnetic Resonance Force Microscopy (MRFM) is a powerful technique to detect a small number of spins that relies on force-detection by an ultrasoft magnetically tipped cantilever and selective magnetic resonance manipulation of the spins. MRFM would greatly benefit from ultralow temperature operation, because of lower thermomechanical noise and increased thermal spin polarization. Here, we demonstrate MRFM operation at temperatures as low as 30 mK, thanks to a recently developed SQUID-based cantilever detection technique which avoids cantilever overheating. In our experiment, we detect dangling bond paramagnetic centers on a silicon surface down to millikelvin temperatures. Fluctuations of such kind of defects are supposedly linked to 1/$f$ magnetic noise and decoherence in SQUIDs as well as in several superconducting and single spin qubits. We find evidence that spin diffusion plays a key role in the low temperature spin dynamics.
\end{abstract}

\maketitle

Magnetic Resonance Force Microscopy (MRFM) is a scanning probe technique based on coupling a soft magnetic tipped cantilever to the spins of a sample, and measuring the tiny force arising from spin manipulation through Magnetic Resonance techniques. The spatial selectivity in MRFM is provided by the strong field gradient generated by the magnetic tip, similar to external field gradients used in conventional MRI. MRFM is nowadays considered one of the most viable routes towards three-dimensional imaging of biomolecules or nanostructures with atomic resolution. While this ambitious goal has not been reached yet, several milestones have already been demonstrated, including mechanical detection of a single electron spin \cite{rugar1} and 3D nuclear spin imaging of a virus with a few nanometers resolution \cite{rugar2}.

Present limiting factors on MRFM resolution are the thermal-noise-limited force sensitivity of the cantilever, and a surface-induced frequency and force noise, which is poorly understood \cite{marohn1}. Besides a better characterization of the latter effect, further improvements require softer and less dissipative mechanical sensors and if possible reduction of operating temperature. So far, MRFM has been demonstrated at temperatures as low as 280 mK \cite{mamin}. Cooling below 100 mK would both improve the force sensitivity and increase the Boltzmann spin polarization, thereby increasing the signal. The last consideration doesn't hold for small nuclear spin ensembles, for which statistical polarization becomes dominant for small spin ensemble \cite{rugar3}, unless temperatures below 1 mK can be achieved.
\begin{figure}[!ht]
\includegraphics[width=86mm]{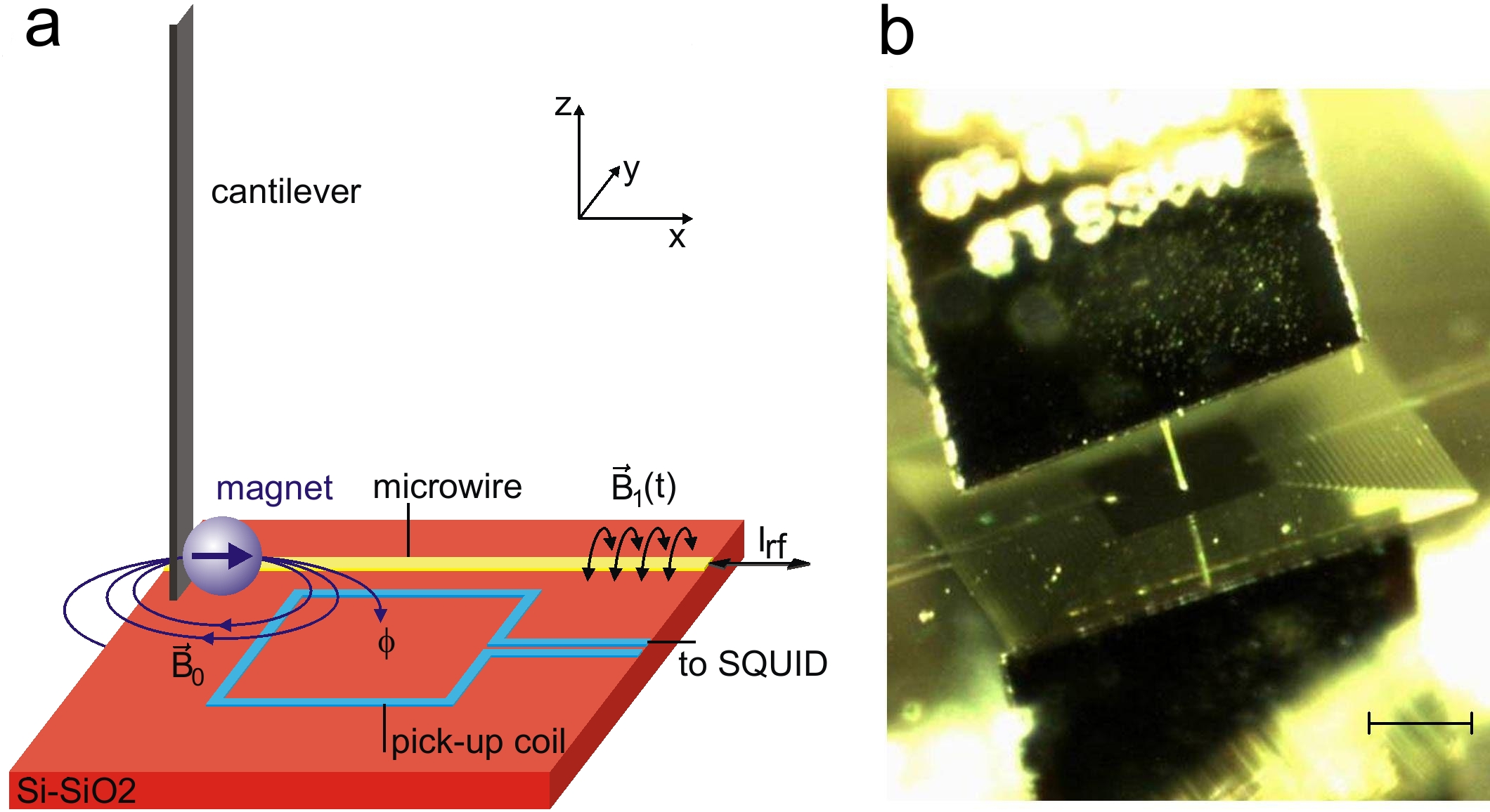}
\caption{\textbf{Experimental scheme.} (a) Schematic of the experimental setup. An ultrasoft cantilever (stiffness $k \approx 10^{-4}$ N/m) with a ferromagnetic particle, the magnet, attached on its end (magnetic moment $\mu \approx 10^{-11}$ A/m) is approached in a vertical configuration to the surface of the detection chip. A superconducting niobium pick-up coil deposited on the chip surface is connected to a dc SQUID. The pick-up coil detects the motion of the magnet through the position-dependent magnetic flux $\phi$ coupled by the magnet into the coil. The magnet generates a quasistatic field $\vec B_0$ which couples to the sample, constituted by paramagnetic electron spins close to the Si-SiO$_2$ surface of the chip. A rf current $I_{\mathrm{rf}}$ flowing in the superconducting niobium microwire generates the $B_1$ field at microwave frequency which excites the electron spins in the sample. (b) Optical microscope picture of the assembled setup for the first experimental run, with a multiturn pick-up coil and without the microwire. The scale bar is 200 $\mu$m.} \label{apparatus}
\end{figure}

A known issue for ultralow temperature operation is the need for a suitable cantilever detection technique. Interferometric detection is an option, but exceedingly low power would be needed to avoid cantilever overheating, unless advanced dielectric mirrors are used to reduce photon absorption in the mirrors. We have recently demonstrated a SQUID-based detection technique, capable of measuring the thermomechanical fluctuations of a cantilever down to 25 mK \cite{usenko}. The scheme is based on detecting the flux change induced in a pick-up coil by the motion of the magnet attached to the cantilever.
 
In this paper, we demonstrate for the first time that this detection system can be readily integrated in a real MRFM experiment. As a case study, we have mechanically detected paramagnetic centers located close to the surface of the chip supporting the superconducting pick-up coil, and subsequently we have performed a MRFM saturation-recovery experiment. Paramagnetic centers appear to be almost ubiquitous on thin film surfaces and interfaces \cite{moler}, with typical density of the order of $0.1 - 1$ nm$^{-2}$, and have been recently subject of various studies, because they are possibly related to the 1/$f$ magnetic noise observed in SQUIDs at millikelvin temperatures \cite{wellstood,ptb,mcdermott1,mcdermott2}. The same magnetic noise appears to be a major source of dephasing in some superconducting qubits \cite{nakamura, bialczak}, and in single dopant qubits \cite{sousa,paik}. Several different mechanisms for 1/$f$ noise generation have been recently proposed, based on electron hopping \cite{clarke1, clarke2} or spin diffusion \cite{faoro}. Our measurements support a picture in which spin diffusion plays a key role in the dynamics of surface paramagnetic centers at millikelvin temperatures.

\section{Results}

\subsection{Experimental scheme}
A general scheme of our experimental is shown in Fig.~\ref{apparatus}. An ultrasoft microfabricated silicon cantilever with a micron-size spherical magnetized particle attached to its end (from now on, the magnet) is approached vertically to the sample chip surface. The cantilever stiffness $k$ is in the $10^{-4}$ N/m range and the magnet moment $\mu$ is of the order $10^{-11}$ A/m. Assuming the $z$-axis perpendicular to the chip surface, both the magnetic moment of the magnet and the cantilever fundamental mode deflection are oriented along the $x$-axis. The motion of the magnet is detected by a superconducting pick-up coil which is connected to a SQUID amplifier \cite{usenko}. The $xy$ position is aligned at room temperature so that the cantilever will still be several tens of microns away from the superconductor lines, when it approaches the surface. The magnet thus interacts mostly with the oxidized silicon surface of the chip. 

We measured the frequency of the cantilever fundamental mode under different experimental conditions, varying the temperature and the distance from the sample, and exposing the sample to microwave radiation. Far from the surface, the cantilever frequency is around 3 kHz. The quality factor is about $3 \times 10^4$ and is reduced to about $1000$ close the the surface at very low temperature. A detailed report and discussion on the dissipation measurements will be the subject of a future paper.
We present data from two experimental runs with slightly different cantilevers, magnets and samples (see Methods for details).

\subsection{Experiment 1: static frequency shift}

\begin{figure}[!ht]
\includegraphics[width=86mm]{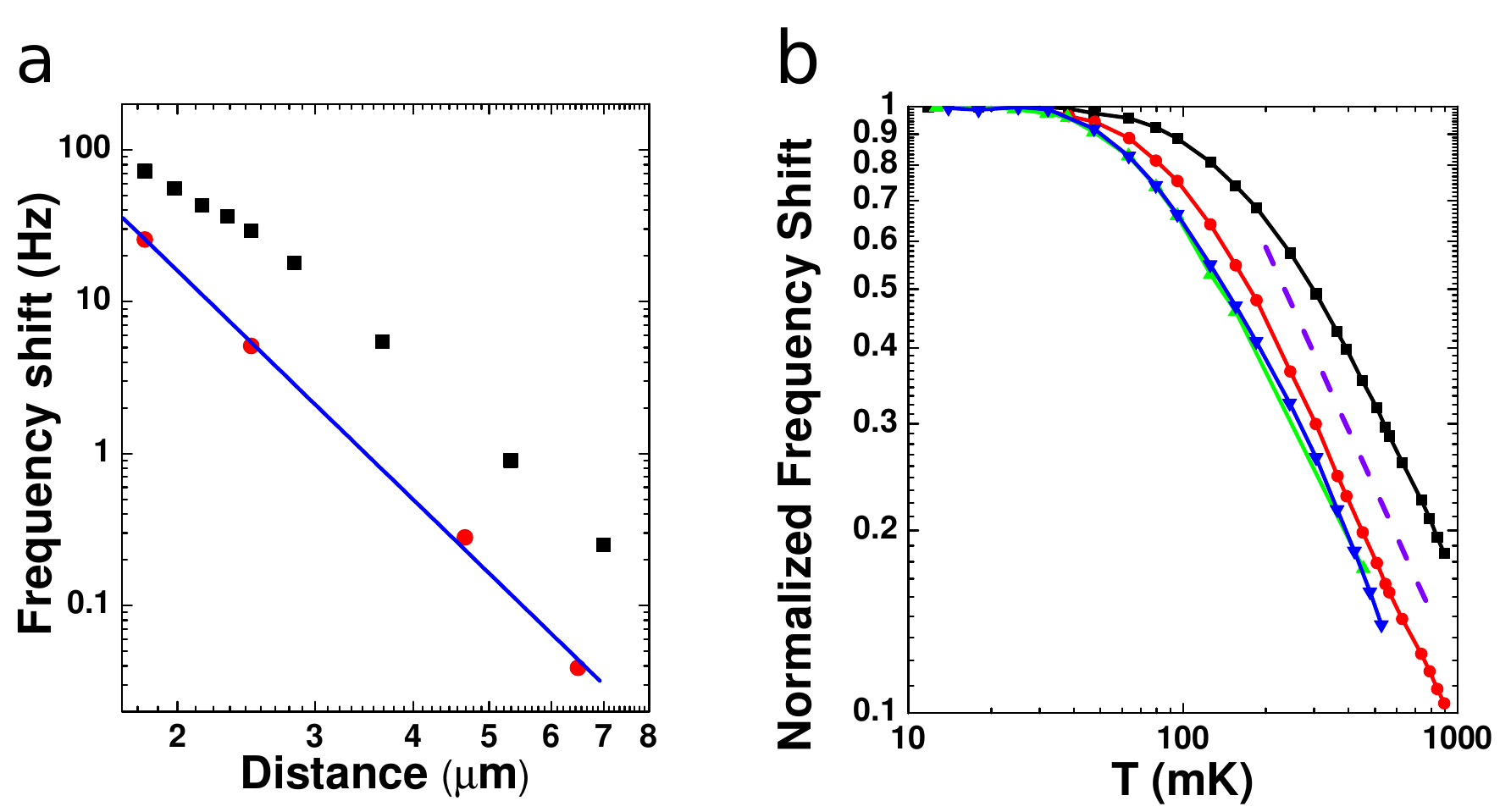}
\caption{\textbf{Static frequency shift as a function of magnet-surface distance and temperature.} (a) Resonant frequency shift $\Delta f_r=f_r-f_{r0}$ in experiment 1 as a function of the distance between magnet center and sample surface, for two representative bath temperatures, $T$=11 mK and $T$=450 mK. Here, $f_{r0}$ is the resonant frequency measured when the magnet is far away from the surface. Straight line shows a log-log linear fit of the 450 mK data, with slope coefficient $n =-5.0 \pm 0.2$. (b) Resonant frequency shift $\Delta f_r$ in experiment 1, as a function of temperature, for 4 different distances between magnet center and sample surface, respectively 6.5 $\mu$m (blue), 4.7 $\mu$m (green), 2.5 $\mu$m (red), 1.8 $mu$m (black). In order to provide an easy comparison of the curves, the data are normalized to the maximum of each curve. The dashed violet line represents the $1/T$ Curie-like behaviour.} \label{static}
\end{figure}

In experiment 1, we accurately measured the resonant frequency of the cantilever fundamental mode $f_r$ as a function of both temperature and distance from the surface, constituted by a $300$ nm thick SiO$_2$ layer on a Si substrate \cite{podt}. We excited the cantilever with a piezoelectric actuator and performed both ringdown measurements and frequency sweeps. Representative experimental results are shown in Fig.~\ref{static}a, where the frequency shift induced by the sample surface is plotted as a function of the distance $d$ between the magnet center and the surface. The power-law dependence and the persistence of a relatively large effect at distances of the order of a few microns rule out short-range van der Waals interactions and point to either electrical or magnetic interactions. Fig.~\ref{static}b shows the frequency shift as a function of temperature for 4 different distances between the magnet center and the surface. The data of each curve are normalized to their maximum value measured at low temperature to provide an easy comparison of the curves. A Curie-like $1/T$ dependence is clearly observed for temperatures higher than 100 mK. This feature is common to all curves, suggesting that the cantilever is indeed coupled to a paramagnetic spin system, for which the polarization increases with decreasing temperature. The low temperature saturation in Fig.~\ref{static}b, together with the anomaly at low distance and low temperature in Fig.~\ref{static}a could indicate either a full polarization of the spins, or a saturation of the effective spin bath temperature.

\begin{figure}[!ht]
\includegraphics[width=86mm]{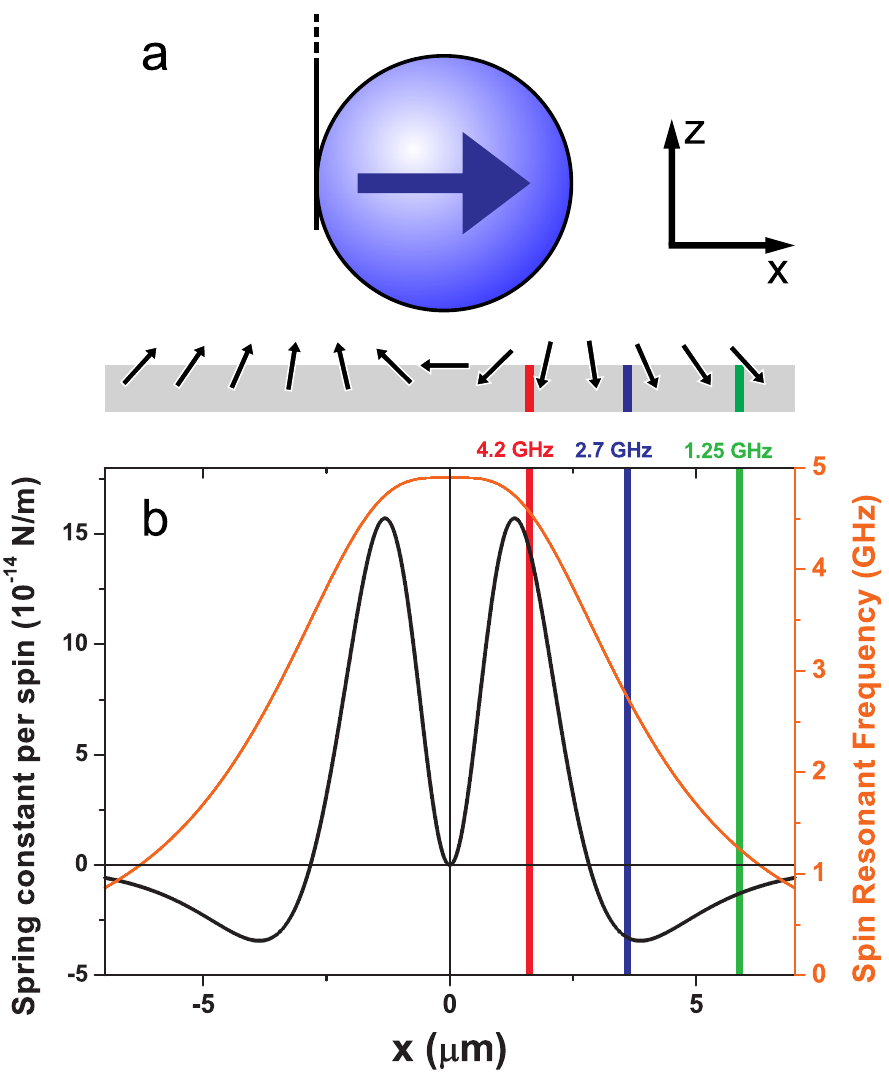}
\caption{\textbf{Model of the magnet coupled to a layer of electron spins.} (a) Scheme of the magnet on the cantilever approaching the surface. It illustrates the fact that surface electron spins are polarized by the magnet dipole field in different directions, and therefore contribute to the cantilever spring constant with both positive and negative sign, depending on the position. (b) Simulated change of the cantilever spring constant $k_s$ induced by a spin located on the surface (black thick line) and magnetic resonance frequency $f=\gamma B_0/2 \pi$ (orange line) as a function of the coordinate $x$, for $y=0$. The parameters correspond to the conditions of experiment 2, with the center of the magnet located at a distance $d=3.6$ $\mu$m from the surface above the origin of the $xy$ plane. The vertical stripes indicate the location of the resonant slices determined by the labelled microwave frequencies.} \label{springconstant}
\end{figure}

In order to account for these observations, we built a simple model to estimate the effect of a layer of paramagnetic spins on the sample surface. Assuming that the longitudinal relaxation time is much longer than the cantilever period, the spring constant induced in the cantilever by an individual electron spin with magnetic moment $\mu_ e$ is given by  $k_s=-\mu_e P B_{xx}$, where $P=\mathrm{tanh}\left(\mu_e B_0 / k_B T \right)$ is the Boltzmann polarization factor and $B_{xx}$ is the second derivative of the magnet field $B_0$ with respect to magnet position. As illustrated in Fig.~\ref{springconstant}, $k_s$ is positive for spins close to the magnet and negative for spins far away from the magnet. Integration over a uniform surface layer with spin density $\sigma$ gives the total spring constant change, from which the frequency shift can be calculated. 

According to the model, we expect a dependence on distance as $d^n$, with $-6<n<-5$ slightly dependent on distance and temperature. Experimental data of Fig.~\ref{static}a are consistent with this prediction. Furthermore the model predicts the observed high temperature $1/T$ dependence in Fig.~\ref{static}b. A fit of the data in the $1/T$ region yields a density of $\left(0.6 \pm 0.1 \right)$ spins/nm$^2$. This figure is reasonably close to the estimates of paramagnetic defects densities in different thin film structures using other techniques \cite{moler,mcdermott1}, in particular it is consistent with recent estimates of the density of dangling bond paramagnetic centers, so called $P_b$ centers, on Si-SiO$_2$ interfaces using single dopant techniques \cite{sousa}. As we are probing a Si-SiO$_2$ interface, the magnet is likely coupled to this latter kind of defects. In principle, electron spins or nuclear spins in the bulk could be coupled to the magnet as well, but for an undoped silicon substrate their effect has been estimated to be at least 2 orders of magnitude smaller.

It is worth mentioning that several past studies on non-contact interaction between a cantilever and a surface, focusing mostly on non-contact friction, found effects due to an interaction of electrical nature, rather than magnetic \cite{stipe, marohn3}. However, non-contact electrical effects are usually found to decrease with decreasing temperature and show a weaker power-law dependence on distance ($n \approx 1-3$) than that we have measured. Therefore, our claim that the measured frequency shift is due to a magnetic interaction with a layer of spins is not in contradiction with previous works. In fact, our measurements are performed at much lower temperature, and we use a relatively large magnet, so that magnetic effects are strongly enhanced.

\subsection{Experiment 2: microwave-induced frequency shift}

In experiment 2 we have applied magnetic resonance pulses to selectively address different spin subsets. In this experiment we used a different detection coil chip, fabricated in a different laboratory with different fabrication protocols. A frequency shift measurement as a function of temperature has been performed as in experiment 1, for a fixed distance $d=3.6$ $\mu$m, which yields a spin density $\sigma =(0.20 \pm 0.05)$ spins/nm$^2$ slightly lower  compared to experiment 1. Spin manipulation was enabled by a niobium microwire, 15 $\mu$m wide, deposited at 50 $\mu$m from the detection coil, through which we could deliver an effective microwave power at low temperature up to $-17$ dBm in a frequency range up to 4 GHz. In this experiment, the magnet on the cantilever was located at a relatively large distance of $70$ $\mu$m from the wire. 

The microwave field $B_1\cong 1$ $\mu$T generated in the vicinity of the magnet was used to saturate the polarized spins, using the so called CERMIT saturation-recovery protocol \cite{marohn2}. The microwave power is switched on for a given time, of the order of the longitudinal relaxation time $T_1$, ranging from a few tens of milliseconds up to a few seconds. The microwave frequency $f_{\mathrm{rf}}= \omega_{\mathrm{rf}}/2 \pi$ defines the resonant slice of spins satisfying the resonant condition $\omega_{\mathrm{rf}}=\gamma B_0$, where $\gamma$ is the gyromagnetic ratio, within a bandwidth determined by the homogeneous broadening. The microwave pulse causes saturation and hence suppression of the polarization of the resonant spins. Subsequently the microwave power is switched off and the thermal equilibrium polarization is allowed to recover. According to standard Bloch equations, the timescale of the polarization suppression process is of the order of $T_1$ in the undersaturated regime, and  $1/ \left( \gamma^2 B_1^2 T_2 \right)$ in the saturation regime, where $T_2$ is the transverse relaxation time, while the timescale of the recovery process is given by $T_1$. We note that in our scheme $B_1$ is parallel to the $z$ axis, and is not purely transverse to the inhomogeneous dipole field $B_0$ generated by the magnet, as in conventional magnetic resonance. While this would be clearly an issue in a pulsed magnetic resonance scheme, it is not a major concern for a saturation experiment. All spins in the resonant slice will be saturated, except a small fraction of them which feel an almost vertical magnetic field $B_0$ field.

To monitor the frequency shift during the saturation-recovery transients, the cantilever is self-oscillated in a phase-locked loop with a 50 Hz bandwidth. We maintained a constant cantilever distance $d=3.6$ $\mu$m and therefore a constant field $B_0$ throughout the following measurements, while we could address different resonant slices by varying the microwave frequency.

\begin{figure}[!ht]
\includegraphics[width=86mm]{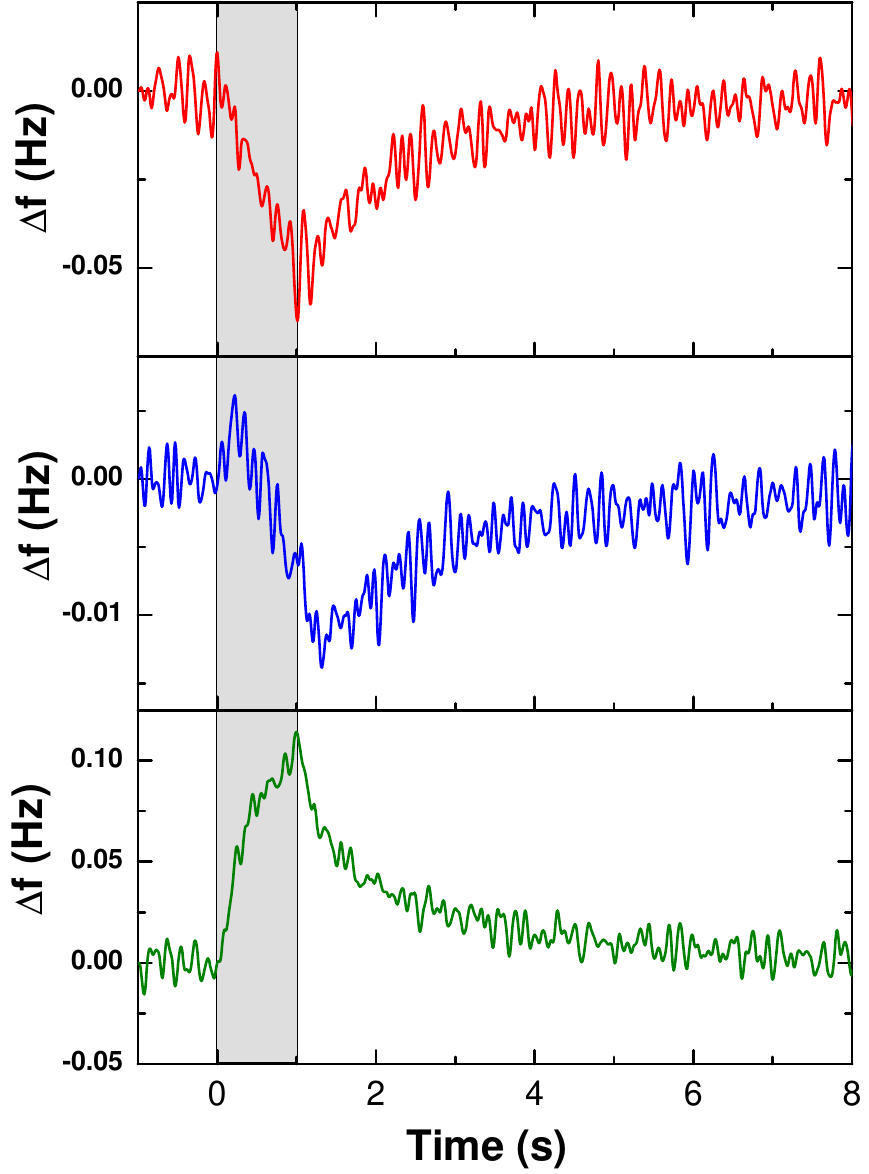}
\caption{\textbf{Microwave-induced frequency shift: saturation and recovery.} Cantilever frequency shift as function of time for 3 different frequencies of the applied microwave pulses, respectively from the top to the bottom 4.2 GHz (red), 2.7 GHz (blue), and 1.25 GHz (green). Here, a microwave power of -23 dBm (corresponding to a field $\approx 1$ $\mu$T) is switched on at $t=0$ and switched off at $t=1$ s. The shaded region indicates the time during which the microwave power is on. At 1.25 GHz the microwave signal induces a positive transient with non-exponential features, while the 4.2 GHz pulse induces a negative exponential transient. At an intermediate frequency of 2.7 GHz a double transient with opposite sign is observed both when switching on and off the microwave.} \label{pulse}
\end{figure}

The resulting saturation-recovery curves for different microwave frequencies $f_{\mathrm{rf}}$ and a nominal power of $-23$ dBm are shown in Fig.~\ref{pulse}. The measurements were done at a bath temperature of 30 mK.
It is apparent that addressing different resonant spins yields qualitatively different behaviours. For $f_\mathrm{rf}=1.25$ GHz the frequency shift is positive, and both transients are slightly non-exponential, with a faster initial behaviour followed by a slower relaxation. For $f_\mathrm{rf}=4.2$ GHz we find a negative signal and exponential relaxation. The physical origin of the sign change is easily understood by inspection of Fig.~\ref{springconstant}. Frequencies around 4 GHz select high field spins close to the magnet, which give a positive contribution to the frequency shift, while frequencies around 1 GHz select low field spins far away from the magnet, giving a negative contribution. Temporarily removing the polarization of these spins gives the signal with the signs we have observed. Finally, at an intermediate frequency of 2.7 GHz we observe a mixed behaviour, with an initially positive transient, which then develops into a net negative frequency shift. In general, the recovery relaxation is characterized by a slow time constant of the order of 1-2 seconds, slightly dependent on the resonant slice.

\section{Discussion}

The observed behaviour at intermediate frequency can hardly be explained in terms of a simple Bloch equation model. Instead, we interpret these data as evidence of spin diffusion mediated by flip-flop processes, which spreads the local suppression of polarization throughout the sample. Spin diffusion also explains the non-exponential behavior at 1.25 GHz. We have checked the faster saturation transient at 1.25 GHz for shorter microwave pulses, finding that the amplitude of the fast frequency response is roughly independent of the microwave power for powers higher than -23 dBm. This rules out direct spurious crosstalk and suggests that the fast response is instead related with the saturation of local spins, followed by a slower diffusion to non-resonant spins.

\begin{figure}[!ht]
\includegraphics[width=86mm]{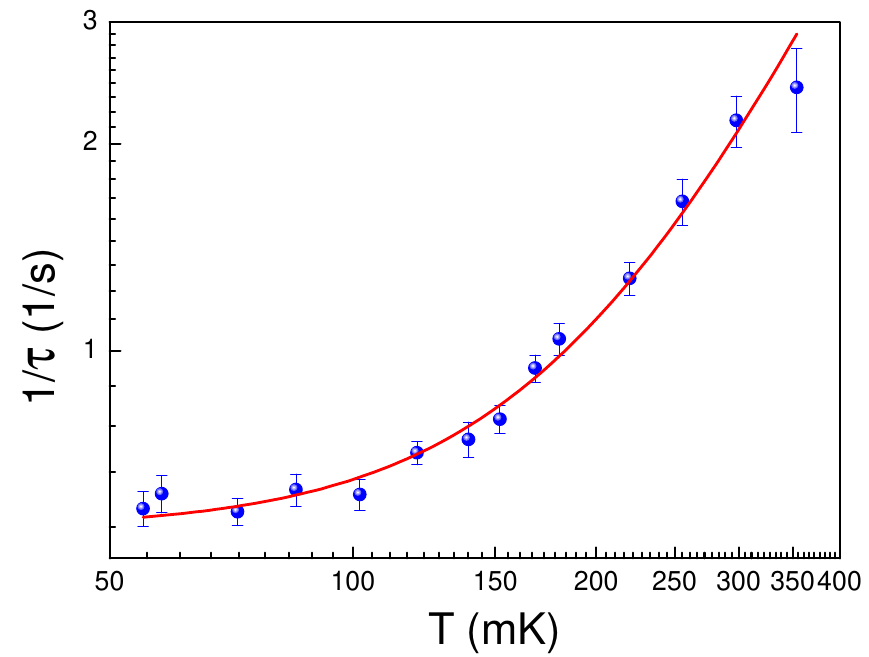}
\caption{\textbf{Recovery relaxation rate as a function of temperature.} Relaxation rate of the frequency-shift signal in the recovery transient (see Fig.~\ref{pulse}) $1/\tau$ as a function of the bath temperature $T$. The solid line represents the best fit of the data with the function $A+B T^n$ , which gives for the exponent of the temperature dependent term the value $n=2.5 \pm 0.2$.} \label{relaxation}
\end{figure}

We now discuss whether a spin diffusion mechanism is expected to be significant in our experiment. A theory of spin diffusion in an inhomogeneous field was developed in the framework of NMR experiments \cite{redfield}. A key point is that a flip-flop process between two spins in the presence of a magnetic field gradient is not energy-conserving, so the diffusion requires that the spin dipolar bath takes up the excess energy.
An important consequence is that diffusion can be quenched by sufficiently large gradients \cite{rugar4}. Application of the model to real situations is generally tricky \cite{meier}, requiring large simulations, but some basic features can be estimated by simple arguments. The diffusion process is controlled by the diffusion constant $D=W a^2$. Here, $a=\sigma^{-1/2}$ is the mean distance between neighbouring spins, and $W$ is the rate of a flip-flop process between two spins. $W$ can be estimated as $W\approx \delta  /30$, with the resonant line width $\delta$ given approximately by \cite{abragam}:
\begin{equation}
  \delta =  3.8 \frac{\mu_0 }{4 \pi} \frac{\hbar \gamma^2}{a^3}.
\end{equation}
For our spin density, we estimate a diffusion constant $D\approx 20$ $\mu$m$^2$/s. A local perturbation of the polarization from its equilibrium value will diffuse across the sample to an approximate diffusion length $L_D=\left( D T_1 \right)^{1/2}$. If we assume the measured recovery relaxation time $\tau \approx 1$ s as an approximate estimation of $T_1$, the corresponding diffusion length becomes $L_D \approx 4.5$ $\mu$m, which is of the order of the characteristic length of the spin system probed by the magnet. This means that spin diffusion can indeed spread the polarization transient throughout the sample. 
A simplified model supporting this picture qualitatively is described in the Supplementary Note, and relevant numerical simulations are presented in Supplementary Figures S1 and S2.
On the other hand, if $T_1$ were much shorter, the diffusion length would be suppressed as direct or indirect spin-phonon relaxation would become dominant. As we expect a temperature dependence of $T_1$, we measured the relaxation time $\tau$ in the recovery transient as a function of temperature (Fig.~\ref{relaxation}). We find that the non-exponential features of the transient are suppressed at higher temperatures and the relaxation rate $1/\tau$ is increasing approximately according to a power law. A fit of the experimental data with the function $A+B T^n$ gives for the exponent the value $n=2.5 \pm 0.2$. Here, a finite value of $A$ accounts for a possible saturation of the spin equilibrium temperature. The temperature dependence of $1/\tau$ is roughly comparable with previous EPR studies on $P_b$ centers \cite{askew}, which have determined $1/T_1 \approx 10$ s$^{-1}$  at 400 mK and a power-law exponent $2.3<n<3.5$.

Because spin diffusion is spreading polarization across the sample, it is hard to define the number of spins generating the frequency shift signals shown in Fig.~\ref{pulse}. However, we can try to estimate an effective number of spins at least for the resonant slice at 1.25 GHz, which shows evidence of saturation of the resonant spins. 
Referring to Fig.~\ref{springconstant}, the thickness of the resonant slice can be estimated from the spin resonance linewidth $\delta$ as $\Delta x = \delta / \left| 2 \pi {\partial f /\partial x} \right|$, where $f$ is the spin magnetic resonance frequency (orange line). Approximating the resonant slice with a circular annulus of radius $x$ and thickness $\Delta x$, the total number of spins in the slice becomes $N\approx 2 \pi \sigma x \Delta x $. Each spin contributes with $k_s$ (black line in Fig.~\ref{springconstant}) to the total spring constant change $\Delta k$. The expected magnitude of the microwave-induced cantilever frequency shift, under the assumption that the resonant spins are saturated, is then given by:
\begin{equation}
\Delta f_r  = \frac{1}{2}\frac{{N k_s }}{k}f_r  \approx \pi f_r \frac{{k_s }}{k}\sigma x\Delta x
\end{equation}
For the 1.25 GHz slice we find $\Delta x \approx 43$ nm, $N \approx 3.1 \times 10^5$, and an expected frequency shift $\Delta f_r \approx 45$ mHz. This rough estimation is in reasonable agreement (within a factor 2) with the experimentally measured signal shown in Fig.~\ref{pulse}, in particular with the amplitude of the fast frequency response. This indicates that the microwave-induced frequency shift is consistent with the static frequency shift, from which we have inferred the spin density $\sigma$, giving further support to the hypothesis that the static effect is indeed due to a magnetic interaction with the paramagnetic spins.

We now discuss the possible connections of our findings with the 1/$f$ magnetic noise observed in thin-film devices such as SQUIDs and superconducting qubits. Recent models point to fluctuations of unpaired paramagnetic electron spins as a likely origin of this excess noise. In particular, a model of 1/$f$ noise from spin diffusion in paramagnetic defects located in the superconductor-insulator interface has been proposed \cite{faoro}. In this model, the emergence of 1/$f$ noise is related to long-range low-frequency spin correlations driven by spin diffusion, whose typical scale is determined by the diffusion length. Diffusion constants in the range of $10-100$ $\mu$m$^2$/s, as estimated in our experiment, imply correlation times of seconds on a 10 $\mu$m length scale, which should couple to the typical size of most thin film devices very well. Thus, we suggest that $P_b$ centers in the insulator interface could be partly responsible for the magnetic noise generation as well. This is in contrast with Ref. \cite{faoro}, where spin densities two orders of magnitude smaller were assumed for paramagnetic centers in the insulator, leading to the conclusion that their role is negligible. Another relevant point is the enhancement of spin diffusion by cooling down to a temperature of the order of 100 mK, due to the observed strong dependence of competing relaxation mechanisms on temperature (Fig.~\ref{relaxation}). This is in agreement with the typical feature of magnetic 1/$f$ noise in SQUIDs and qubits, which increases sharply only when the temperature is reduced below 1 K \cite{wellstood, ptb}.

We conclude with the implications of these results for MRFM at ultralow temperature. The interaction of paramagnetic centers in the Si-SiO$_2$ interface with the magnetic tip and with the spins in a sample leads to undesired cantilever damping and shortened spin coherence time, respectively. In order to advance further towards single atom imaging by MRFM, it will be necessary to either develop methods to shield the experiment from the $P_b$ centers, e.g. by applying a background magnetic field in combination with low temperatures to freeze $P_b$ centers dynamics, or to resort to substrates that contain a sufficiently low concentration of paramagnetic defects.

\section{Methods}

\subsection{Experimental apparatus}

In the first experiment, the cantilever was a micromachined IBM-type ultrasoft silicon beam \cite{IBMcantilever} with length, width and thickness respectively 120 $\mu$m, 5 $\mu$m and 100 nm.  The magnet was a 3.3 $\mu$m diameter spherical particle from a commercial neodymium-alloy powder (Magnequench, type MQP-S-11-9) with saturation magnetization $\mu_0 M_r=\left( 1.3 \pm 0.1 \right)$ T \cite{hammel}. The particle is glued to the free end of the cantilever by means of platinum electron beam deposition and is subsequently magnetized in a 5 T field at room temperature. Its magnetic moment is estimated from the volume and the saturation magnetization to be $\mu\approx 2 \times 10^{-11}$ A/m. In the second experiment, the length of the cantilever was 90 $\mu$m and the magnet was 5.2 $\mu$m in diameter, with magnetic moment $\mu \approx 8 \times 10^{-11}$ A/m. 

We also used two different detection-sample silicon chips. The first chip supports a multiturn niobium coil fabricated in a multilayer process, with inner and outer size of respectively $220$ $\mu$m and $660$ $\mu$m, a total number of $44$ turns, and an estimated inductance of $0.6$ $\mu$H. After fabrication of the coil, a $300$ nm thick insulating layer of silicon dioxide is deposited on the top of the chip by rf sputtering \cite{podt}. In the second experiment, we used a different chip, containing a single loop square Nb pick-up coil, with 100 $\mu$m size, and a 5 $\mu$m niobium microwire for microwave excitation. There was no intentional oxide layer in addition to the native oxide.

The flux change in the pick-up coil is measured by a two-stage Superconducting Interference Device (SQUID) amplifier with input inductance $L_i=1.6$ $\mu$H and a constant flux noise of $0.6$ $\mu \phi_0/\sqrt{\mathrm{Hz}}$ below 500 mK \cite{usenko}. In the second experiment we used an intermediate superconducting transformer to match the pickup coil and SQUID inductance. The overall sensitivity of the SQUID-based displacement detector was a few pm/$\sqrt{\mathrm{Hz}}$ in both experiments.

The cantilever was mounted on a piezo-motor, which allows to approach the chip surface with 170 nm steps. 
During the microwave saturation-recovery measurements, a piezo-tube actuator was used to drive the cantilever fundamental mode and a phase-locked loop was used to track the cantilever resonant frequency.
The position of the cantilever was aligned at room temperature so that the magnet approaches the sample on the dielectric in a position, which provides a reasonable coupling between the magnet motion and the coil, while maintaining the magnet more than 20 $\mu$m away from the superconducting lines.
The whole assembly was mounted on a vibration isolation stage, which is thermalized to the mixing chamber of a cryofree dilution refrigerator with a base temperature of 10 mK.

\section{Acknowledgments} 
This work was partially supported by the European Project MicroKelvin and by an ERC starting grant. We thank G. Koning, D. Van der Zalm, R. Kohler, M. Hesselberth and F. Galli for technical help.

\section{Author contributions}
T.H.O., G.W., A.V., O.U. conceived and carried out the experiment, A.V., G.W. performed data analysis and simulations, A.V. wrote the paper, L.S. designed and fabricated the superconducting coil, all authors discussed and commented the results.

\section{Competing financial interests}
The authors declare no competing financial interests.

\end{document}